\documentclass[pdflatex,sn-chicago]{sn-jnl}% Chicago-based Humanities Reference Style

\usepackage{graphicx}%
\usepackage{multirow}%
\usepackage{amsmath,amssymb,amsfonts}%
\usepackage{amsthm}%
\usepackage{mathrsfs}%
\usepackage[title]{appendix}%
\usepackage{xcolor}%
\usepackage{textcomp}%
\usepackage{manyfoot}%
\usepackage{booktabs}%
\usepackage{algorithm}%
\usepackage{algorithmicx}%
\usepackage{algpseudocode}%
\usepackage{listings}%
\usepackage{array}
\usepackage{booktabs}
\usepackage{amsmath} 
\usepackage{multirow}
\usepackage{array}

\begin{document}

\title[Why do people think liberals drink lattes?]{Why do people think liberals drink lattes? How social media afforded self-presentation can shape subjective social sorting}


\author*[1]{\fnm{Samantha C.} \sur{Phillips}}\email{samanthp@cs.cmu.edu}

\author[1]{\fnm{Kathleen M.} \sur{Carley}}\email{kathleen.carley@cs.cmu.edu}

\author[2]{\fnm{Kenneth} \sur{Joseph}}\email{kjoseph@buffalo.edu}

\affil*[1]{\orgdiv{Software and Societal Systems}, \orgname{Carnegie Mellon University}, \orgaddress{\street{4665 Forbes Ave}, \city{Pittsburgh}, \postcode{15213}, \state{PA}, \country{USA}}}

\affil[2]{\orgdiv{Computer Science and Engineering}, \orgname{University at Buffalo}, \orgaddress{\street{12 Capen Hall}, \city{Buffalo}, \postcode{14260}, \state{NY}, \country{USA}}}


\abstract{Social sorting, the alignment of social identities, affiliations, and/or preferences with partisan groups, can increase in-party attachment and decrease out-party tolerance.
We propose that self-presentation afforded by social media profiles fosters \textit{subjective} social sorting by shaping perceptions of alignments between non-political and political identifiers.
Unlike previous work, we evaluate social sorting of naturally occurring, public-facing identifiers in social media profiles selected using a bottom-up approach.
Using a sample of 50 million X users collected five times between 2016 and 2018, we identify users who define themselves politically and generate networks representing simultaneous co-occurrence of identifiers in profiles.
We then systematically measure the alignment of non-political identifiers along political dimensions, revealing alignments that reinforce existing associations, reveal unexpected relationships, and reflect online and offline events.  
We find that while most identifiers bridge political divides, social sorting of identifiers along political lines is occurring to some degree in X profiles.
Our results have implications for understanding the role of social media in facilitating (the perception of) polarization and polarization mitigation strategies such as bridging interventions and algorithms.}




\maketitle

\section{Introduction}\label{sec1}

Mass political polarization in the United States in recent decades can be characterized by two major trends.
First, longitudinally conducted public opinion surveys demonstrate an increase in affective polarization; that is, negative attitudes and distrust between partisan groups (e.g., Democrat, Republican) \cite{iyengar2019origins,abramowitz2016rise}.
Second, American partisan identities are growing increasingly aligned with ideological identities\footnote{\url{https://www.pewresearch.org/politics/2014/06/12/political-polarization-in-the-american-public/}} \cite{mason2018one}, attitudes towards policy issues \cite{dellaposta2020pluralistic,baldassarri2008partisans}, lifestyle preferences \cite{dellaposta2015liberals}, and other social identities (e.g., demographics) \cite{mason2018one,mangum2013racial}.

The latter phenomenon, where partisan identities increasingly align with non-partisan identities\footnote{or, equivalently, vice versa}, is known in the literature as \textit{social sorting} \cite{mason2015disrespectfully,mason2016cross}. Social sorting can exasperate social division (i.e., affective polarization) by reducing the number of cross-cutting lines of conflict \cite{blau2018crosscutting,coser1956functions}. These cross-cutting dimensions of conflict can limit affective polarization because they are capable of lessening perceived differences between groups and weakening identification with a singular dominant identity (e.g., partisan identities).
This limiting of perceived differences across partisan groups, in turn, reduces intolerance and bias towards outgroups \cite{roccas2002social,brewer2005social} and mitigates negative reactions to political messages \cite{mason2016cross}.
For example, even if two people support different presidential candidates, if they also go to the same church or support the same sports team, they can find common ground that alleviates interpersonal conflict.

Actual social sorting---known as \emph{objective} social sorting---refers to the true alignment between groups, and can be estimated using a nationally representative survey (e.g., the American National Election Study) \cite{mason2018one,dellaposta2020pluralistic,dellaposta2015liberals}.
It is thought that objective social sorting should only affect social attitudes if people can correctly identify the alignment between their partisan identity and other ingroups \cite{mason2018one,levendusky2009partisan}.
However, people tend to overestimate the proportion of partisan group members that belong to party-stereotypical groups \cite{ahler2018parties}.

Conversely, perceived social sorting---known as \emph{subjective} social sorting---can induce greater partisan attachment and identity strength regardless of objective alignments \cite{mason2018one}.
Subjective social sorting can contribute to greater perceived (``false'') polarization, in which citizens overestimate social and ideological divisions between groups \cite{westfall2015perceiving}.
Scholars have suggested perceived polarization may be as powerful, if not more, than actual out-group attitudes in driving partisan dislike and distrust \cite{wilson2020polarization,ahler2018parties,enders2019differential}.
The natural question follows: what, or who, shapes perceived polarization of the public? That is, how does subjective social sorting, in particular subjective social sorting that differs from objective sorting, come to be?

Previous work has examined the role of political elites\footnote{definitions of ``political elite'' vary from solely politicians to inclusion of any individual or organization with sizable political influence; we use the latter definition} and mass media in generating partisan (mis)perceptions and contributing to political polarization \cite{banda2018elite,druckman2013elite,wilson2020polarization,levendusky2009partisan}.
Political science theory connecting elite polarization to social sorting proposes a more polarized elite produces clearer, more distinct cues for the public \cite{levendusky2009partisan}.
As voters see the positions and identities signaled by political elites, they associate these signals with all members of the respective political group (i.e., they perceive heightened social sorting), which in turn increases ideological and affective distance between groups, as well as in-party attachment and ideological homogeneity (for people whose identities align).
%As voters see and adopt the positions and identities signaled by political elites, they sort, which in turn increases in-party attachment and ideological homogeneity, as well as ideological and affective distance between groups.

The present work explores a different kind of political elite---partisan social media users---that may be playing an important role in shaping subjective social sorting, and hence perceived polarization, via digital platforms. 
Social media plays a key role in (mis)information exposure and sharing, enabling people to connect and share ideas across geographic bounds.
On these platforms, ordinary people (e.g., not elected officials, celebrities, news agencies) can obtain great and widespread influence in and out of politics \cite{riedl2021rise}. 

A key feature of many of these social media platforms, including  X (formerly Twitter), is allowing users to write a bio: a brief description of the (owner of the) account \cite{madani2023measuring}.
Users can select personal identifiers \cite[any phrase that describes the owner of the social media account: social identities, preferences, activities, and affiliations][]{pathak2021method} to put in their profile \cite[``self-presentation''][]{marwick2011tweet} that they believe are advantageous given their goals and perceived audience.
These identifiers can influence the signals used in profiles of fellow in-group members through self-reinforcing dynamics of homophily and social influence, perpetuating associations between certain identities \cite{luders2022becoming,dellaposta2015liberals,yoder2020phans}.
Furthermore, they can repel out-group members by making between-group differences salient, potentially inspiring countering signals in out-group user profiles \cite{luders2022becoming}.

Critically, repeated exposure to co-occurring identifiers in user profiles can reinforce perceptions of alignment, contributing to subjective social sorting.
Indeed, there is strong evidence that repeated exposure to information (e.g., group attributes) induces belief due to mechanisms like repetition induced cues for truth (e.g., familiarity, fluency) \cite{ouguz2022repeating}.

Taken together, we suggest that social media users that use explicitly political identities (e.g., ``republican'', ``maga'', ``liberal'') to represent themselves, among other identity signals, influence perceptions of alignments between political identities and other groups.
%These users encapsulate more than politicians and partisan media, representing a self-selecting group of political influencers.
In other words, social media may be shaping subjective social sorting with political identities by affording self-presentation through user profiles.

What remains largely unaddressed, however, is an empirical investigation of such users---their prevalence, behaviors (over time), and which identities and/or identifiers they present alongside partisan identities. To this end, the present work begins by addressing the following research question:
\begin{quote}
    \emph{\textbf{RQ1}: How and to what extent do X users publicly define themselves using political identities, and what (if anything) differentiates these accounts from users without political identities in their profile?} 
\end{quote}
To do so, we examined the profiles of a sample of approximately 50 million X users five times between September 2016 and December 2018.
After extracting the most common phrases from bios (used by $\ge 1000$ users), we used a mixed-methods approach to 1) identify phrases in bios that represented identities, and 2) determine which of those signaled a political identity.

After identifying self-defined political users, we turned our attention to the question of how subjective social sorting manifests in X bios via the ways in which partisan users opt to self-present in their bios. 
Specifically, we ask:
\begin{quote}
    \emph{\textbf{RQ2}: Is there evidence of subjective social sorting of personal identifiers (social identities, preferences, activities, and affiliations) and political identities in X user profiles?  If so, how?}
\end{quote}
To address this question, we use semantic network analysis to quantify alignment between identifiers and political identities.
We examine alignments on two levels of granularity---individual identifiers and categories of identifiers---to provide a more comprehensive understanding of how social sorting manifests in X user profiles.
In analyzing categories of identifiers, we want to determine if there are high-level patterns of identifier use by domain across political groups.
Specifically, we ask:
\begin{quote}
    \emph{\textbf{RQ2a}: Do different types of identifiers reflect more bias towards and/or division between political groups?  If so, how?}
\end{quote}
To evaluate the extent to which different types of identifiers (e.g., religion, sexuality/gender, family) differentiate users holding political identities, we  manually annotated each identifier into zero, one or two of 21 non-political categories.  
For each category and time period, we calculated the average and variance of differences in alignment between groups calculated using semantic network analysis.

Furthermore, to emphasize the utility of our approach in assessing not only subjective social sorting, but how it changes over time, we analyze subjective social sorting as observed through self-presentation in bios in the context of a major shift in American political parties: the catalyzation of an anti-establishment conservative movement (largely) by Donald Trump \cite{hopkins2022trump,oliver2016rise}.
Specifically, we ask:
\begin{quote}
    \emph{\textbf{RQ2b}: How does social sorting evolve between September 2016 (pre-Trump presidency) and December 2018 (mid-Trump presidency)?}
\end{quote}
In the context of the shift in party identification that occurred throughout this period, left-right extremity and partisan tribalism cannot fully explain the rise in partisan animosity \cite{uscinski2021american}.  
Another dimension that has been argued and shown to be crucial in explaining increases in political violence and toxic rhetoric, as well as mainstream support for conspiracy theories and populism, is anti-establishment attitudes \cite{uscinski2021american,oliver2016rise}.  
Therefore, we examine subjective social sorting on X along two dimensions: left/right and pro-/anti-establishment. 

In addressing the four research questions, the present work makes the following contributions to the literature:
\begin{itemize}
    \item We show that political identities were exceedingly rare in X bios during a critical time in American politics.  
    We also show, however, that X users increasingly defined themselves politically during this period (in particular, using ``MAGA'' and ``Trump'') and tend to be more active and have more followers than the average user.
    %, but increase by 18\% between September 2016 and December 2018, particularly due to additions of ``MAGA'' and ``Trump''.  In addition, we compare activity and popularity of users with and without political identities in their profile.
    \item We provide empirical evidence of strong alignments between approximately 9.2\% of non-political identifiers and political identities in X user profiles.
    \begin{itemize}
        \item We highlight alignments that reinforce existing relationships, reveal unexpected associations, and reflect offline and online events.
        \item We characterize the extent to which different types of identifiers differentiate users along each dimension, indicating certain domains exasperate divisions while others bridge divides.
        \item We demonstrate how temporal changes in alignment reflect broader shifts in party identification as holders of anti-establishment orientations joined and emerged from the Republican party.  
    \end{itemize}
    \item We present (and, upon acceptance, will make available to researchers) a novel dataset of the profile bios for over 50 million English-language Twitter users, captured at three month intervals over two years from 2016-2018.
\end{itemize}

\section{Related Works}

Political polarization has been studied extensively on digital platforms through a wide range of operational forms.
Previous work analyzed the segregation of communication (e.g., retweets, @-mentions) and content use (e.g., hashtags) networks between partisan users to shed light on the extent to which these like-minded groups resemble echo chambers  \cite{conover2011political,guerra2013measure,garimella2017long,phillips2023high}.
Other scholars examined news sharing and consumption along partisan lines to investigate the diversity of biased media within like-minded communities  \cite{weld2021political} and the degree to which partisan preferences drive online news selection \cite{garimella2021political}.
It is clear from these studies that polarization, to some extent, is reflected and/or enhanced by online platforms.
However, they only represent a few of many ways to think about polarization.

Social scientists have demonstrated growing alignments of identities and beliefs along partisan lines (i.e., objective social sorting) using longitudinally rich and nationally representative surveys \cite{mason2016cross,dellaposta2015liberals,dellaposta2020pluralistic,mason2018one}.  
Unlike previous research, we measure social sorting of naturally occurring, public facing identifiers in social media profiles of self-selected political influencers.

Beyond increasing strength of alignments, there is evidence that an increasing number of attitudes and preferences are aligning with partisanship \cite{dellaposta2020pluralistic}.
Crucially, any increased breadth of alignments would likely not be detected if researchers use a pre-selected list identities and beliefs of interest. 
Therefore, we do not pre-select identities or beliefs of interest, which typically represent existing political debates.
Rather, we use a bottom-up approach to identify the complete set of personal identifiers (used by at least 1000 users in the total sample) presented alongside political identities in user profiles.
The dimension(s) in which we assess social sorting (left/right, pro-/anti-establishment) is also flexible, in contrast with survey data.

Self-identifying partisan X users have been studied previously in terms of their prevalence \cite{rogers2021using}, demographics and activity levels \cite{barbera2015tweeting,conover2012partisan}, and even presentation of cultural and lifestyle preferences \cite{essig2024partisan}.
Our sample of X users is larger and less heavily biased towards highly active users than previous analyses of political users that used a random 1\% sample of public facing tweets \cite{rogers2021using,essig2024partisan} or keyword searches \cite{barbera2015tweeting}, meaning our findings have higher external validity.
We also have the added benefit of temporal data, so we can assess changes over time.

Instead of using pre-defined stopword lists or frequency to select terms of interest extracted from user bios, we go a step further to select personal identifiers based on theoretical significance.
In addition, our alignment measure incorporates the underlying distribution of identifiers by including Bayesian priors.
In this way, we assess the extent to which each identifier is used differently by users defining themselves with left vs. right and pro- vs. anti-establishment political identities, while mitigating effects of sampling variability and over-fitting.

\section{Data}\label{sec2}
The data used in this work is the meta data of a sample of 308,141,986 X users, representing the most active users in 2010-2015, collected via Twitter API in September 2016, June 2017, December 2017, June 2018, and December 2018.  Approximately 30M of these users have an empty bio, which we filtered out.  In addition, we only included English-language bios because of our interest in American politics.  We end with approx. 50 million users in each time period.

\section{Methods}\label{sec3}

\subsection{Identifier extraction}
For each user in each time period, we extracted phrases from bios using the method developed by \citet{pathak2021method}. Their work develops the concept of a ``personal identifier''---roughly, a phrase that signals identity---and proposes a simple method to extract personal identifiers from text. However, because their method generates noisy phrases, we opt to develop methods to post-process the output of their model, as well as methods to subsample to only frequent phrases.

Specifically, we first labeled 2,271 phrases used by at least 5,000 users as personal identifiers or not, obtaining 1,177 identifiers. To do so, two authors labelled a random sample of 200 phrases, obtaining a Krippendorf's alpha of 0.8005.  
One author then labelled the remaining 2,271 phrases.
To facilitate the labeling process for additional phrases, we then trained classical machine learning models using a Sentence-BERT model fine-tuned using X bios to generate embeddings  \cite{madani2023measuring}; see Appendix. 
We use this model as a filter for 11,141 phrases used by at least 1,000 users in one time period, obtaining 5,143.  
Finally, we conducted a last manual filtering phase to remove non-personal identifiers.  This process left us with a set of 3,670 identifiers in bios that we analyzed.

\subsection{Identifier categorization}
Of the selected identifiers, we labeled those that explicitly reference partisan groups; see Table \ref{table1}.  All authors developed and agreed on the labels by reviewing the related literature \cite{uscinski2021american,oliver2016rise} and media\footnote{\url{https://www.cnn.com/2020/03/03/politics/democratic-establishment-reaction-bernie-sanders/index.html}}\footnote{\url{https://www.politico.com/magazine/story/2019/11/07/media-biased-against-elizabeth-warren-229907/}}.  
We broadly use ``political identities'' to refer to both partisan and ideological identities that distinctly present one side of at least dimension.  We group direct support for clearly partisan politicians in with partisan group identities for the purposes of this work.

\begin{table}[t]
\centering
%\resizebox{.95\columnwidth}{!}{
\begin{tabular}{l|p{3cm}|p{3cm}}
& Pro-establishment & Anti-establishment \\ 
\hline 
Left & democrat, imwithher, liberal, stillwithher, Obama, Hillary & socialist, progressive, feelthebern, Warren  \\
\hline 
Right & (fiscal, consitutional) conservative, republican, gop & MAGA, trump2016,  (Donald) Trump, libertarian, Trumptrain, make america great again \\
\end{tabular}
\caption{Political identities.  Other partisan identifiers (not inherently pro- or anti-establishment): nevertrump, neverhillary, notmypresident, uniteblue, lefty.}
\label{table1}
\end{table}
For \textit{RQ2a}, we evaluate identifiers categorized as: culture, nationality/location/ethnicity/race, occupation/industry, sports, school/subjects,  technology/gaming/cars, (social) media, horoscope, travel, religion, outdoors/nature/animals, business/finance, activism, disability/(mental) illness, sex, military, age, family, values, relationship status, sexuality/gender.  
A clarifying point: culture refers to music, art, food, fashion, and other cultural preferences and roles. 
The categories were selected based on previous research on self-presentation on social media \cite{yoder2020phans,pathak2021method}.

Two authors labelled a random sample of 420 identifiers and obtained a Krippendorff's alpha $>0.65$ for each categories (mean:  0.856, standard deviation: 0.098); see Table D3
%\ref{tab:category_counts}
in the Appendix for further details. 
A single author labelled the category(-ies) of the remaining identifiers that co-occurred at least once with a political identifier.

\subsection{Alignment of identifiers with political identities}

For \textit{RQ2(a,b)}, we generate a co-occurrence network where the nodes are identifiers (n=3670) and edges are weighted by the number of users who have both identifiers in their bio simultaneously for each time period.  
Table \ref{tab:co_occur} in the Appendix describes summary statistics for each network.

We then measure the alignment of each identifier along the specified dimension (e.g., left vs. right) by calculating the difference in weighted log-odds ratios with informative Dirichlet priors between the two groups \cite{monroe2008fightin} using the tidylo R package\footnote{\url{https://github.com/juliasilge/tidylo}}. 
That is, the usage difference of identifier $w$ between two groups $i$ and $j$ at time $k$ is calculated by
\begin{align}
    \delta_{kw}^{(i-j)} &= log\left[\frac{y_{kw}^{(i)} + \alpha_{kw}^{(i)}}{n_k^{(i)} + \alpha_{k0}^{(i)} - y_{kw}^{(i)} - \alpha_{kw}^{(i)}} \right] \\
    &- log\left[\frac{y_{kw}^{(j)} + \alpha_{kw}^{(j)}}{n_k^{(j)} + \alpha_{k0}^{(j)} - y_{kw}^{(j)} - \alpha_{kw}^{(j)}} \right]
\end{align}
where $y_{kw}^{(i)}$ is the number of users that used an identifier in group $i$ (e.g., left identifiers) and identifier $w$ in time period $k$, and $n_k^{(i)}$ is the number of users that used an identifier in group $i$.  The informative prior is $\alpha_{kw}^{(i)} = $ \textbf{y} $\cdot \frac{\alpha_0}{n}$, where $\alpha_{0} = \sum_w \alpha_w$, $\alpha_{k0}^{(i)} = \sum_w \alpha_w^{(i)}$, \textbf{y} $\sim$ Multinomial($n, \boldsymbol\pi$) are the counts in the entire corpus, $n = \sum_w y_w$ and $\boldsymbol\pi$ is a $W$-vector of multinomial probabilities.

We can then divide by the standard deviation of $\delta_{kw}^{(i-j)}$ to obtain z-scores of the log-odds-ratios:
\begin{equation}
    \zeta_{kw}^{(i-j)} = \hat{\delta}_{kw}^{(i-j)} / \sqrt{\sigma^2 (\hat{\delta}_{kw}^{(i-j)})}
\end{equation}

We then measure the bias and polarization of identifiers by calculating the mean and standard deviation of $\zeta_{kw}^{(i-j)}$, respectively, for each category of identifiers and time period.  
Specifically, the mean difference represents how biased that category of identifiers (e.g., military, family, gender/sexuality) is towards one side of the given dimension (e.g., left - right).  
%These biases indicate the types of political identities that use identifiers in that category more on average.
The standard deviation quantifies the variance of identifiers across partisan groups.
A larger standard deviation means there is more division in use of identifiers in that category by users on either side of the dimension.
%It complements the information provided by the average to indicate if 1) each side uses different identifiers in that category or 2) only one side is using that type of identifier.  
We adopt these interpretations from previous work that used the mean differences as a measure of partisanship and standard deviation of differences as a polarization measure itself \cite{monroe2008fightin}.  

%Labeled identifiers, the personal identifier classifier, and co-occurrence networks between identifiers will be released upon acceptance.

\subsection{Ethics}
We only collected public-facing data in accordance with Twitter Terms of Service at the time and did not interact with users in any way.  
In this work, we only use derivatives of the collected data such that no personal identifying information (e.g., sequence of identifiers in individual bios) can be used to reveal the users in our study  
(available via OSF\footnote{OSF LINK to be added}).
%The original data is stored on secure password-protected servers.  
The anonymized original data, that contains information that \textit{could} identify an individual user in some cases, is available upon request to researchers who commit to only using the data for non-commercial, research purposes.  See Appendix for discussion of FAIR principles compliance.

\section{Results}\label{sec4}

\subsection{Overview of political identification on X between 2016 and 2018}

%In this section, we provide important context for our investigation into subjective social sorting along partisan and ideological lines by describing how and the extent to which X users explicitly define themselves politically. In addition, we compare activity level and influence of users with and without political identities in their profile to motivate analysis of this particular set of political influencers.

\begin{figure*}[t]
\centering
\includegraphics[width=\textwidth]{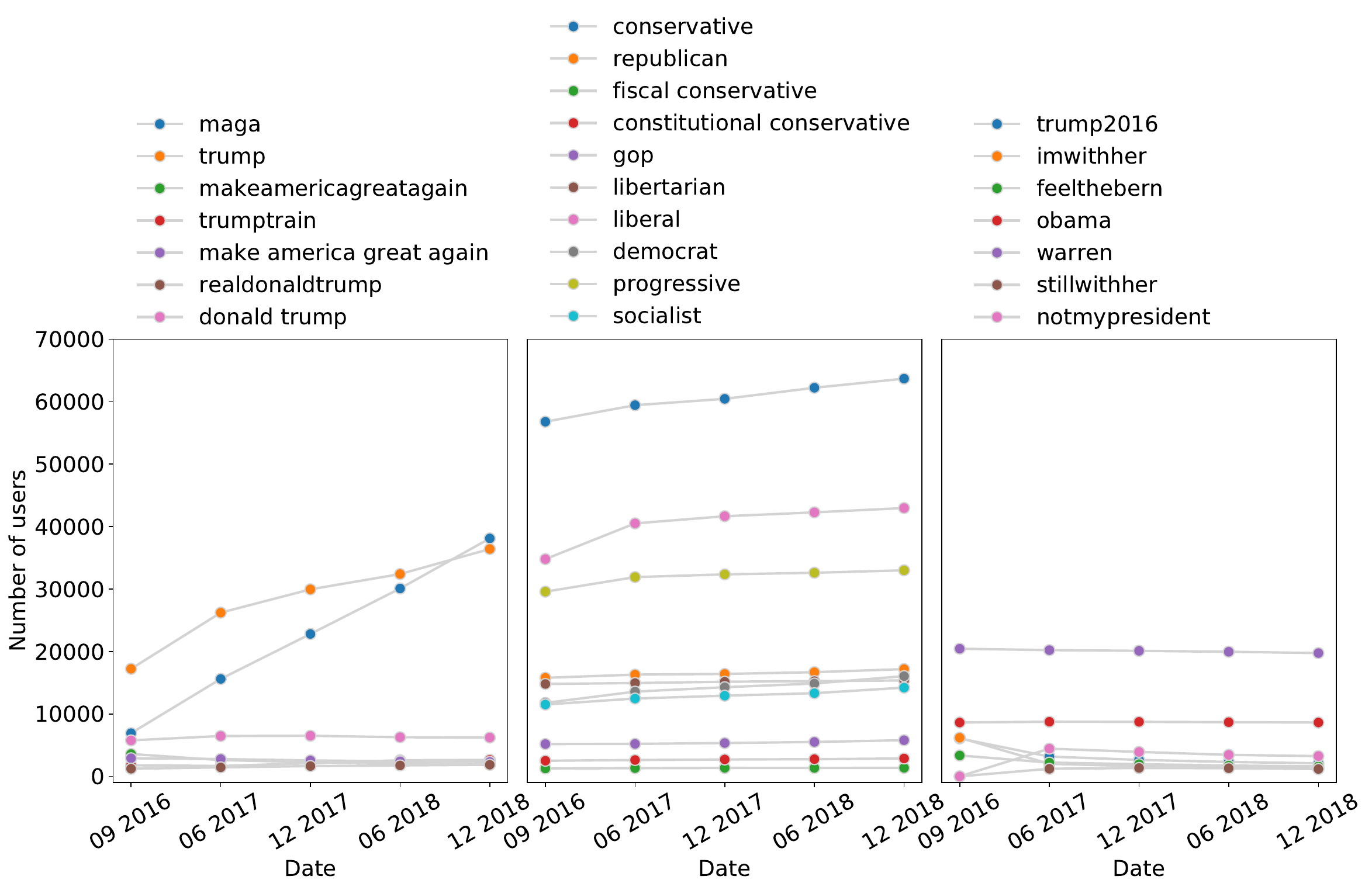}
\caption{Number of users with political identities in their bio per time period. Division of identifiers selected to maximize clarity.}.
\label{fig1}
\end{figure*}

Figure \ref{fig1} contains the number of user profiles that contain each political identity from Table \ref{table1} over time.  
Most notably, there was a striking and consistent increase of ``MAGA'' and ``Trump'' in bios throughout 2017 and 2018, where ``MAGA’’ increased by 450\% (6,918 to 38,105) and ``Trump’’ increased by 85\% (22,992 to 42,627).  
Overall, the number of users holding right anti-establishment identities increased by 56\% (70,129 to 109,696) over this period; see Table \ref{tab:counts} in the Appendix.  
Even before Trump was elected as president, other politicians were simply not referenced in bios to the same extent (e.g., Hillary Clinton, Elizabeth Warren), echoing tweet-level analyses of the 2016 election X discourse \cite{oh2017trump}.  
This reflects Trump's relative success in strategically using social media to disseminate his right-wing populist ideas directly to his followers and draw engagement \cite{kreis2017tweet}.

We also see an increase, albeit less dramatic, in the number of users publicly holding pro-establishment right identities by 12\% (80,654 to 90,318).  
The largest total increase was ``Conservative’’ by 12\% (60,564 to 67,922), followed by ``Republican’’ by 9\% (15,787 to 17,182) and ``GOP'' by 12\% (5,181 to 5,793). 
In addition, we find pro- and anti-establishment right identities co-occur substantially more between September 2016 and December 2018.  
Specifically, the number of profiles that contain ``Conservative’’, ``Republican'', or ``GOP''  and at least one anti-establishment right identity increased by 142\% (6,503 to 15,724), 67\% (1,842 to 3,081), and 103\% (587 to 1,193), respectively.  
These changes in co-occurrence demonstrate the fundamental shift of party identification that occurred, as holders of anti-establishment orientations joined and/or emerged from the Republican party \cite{hopkins2022trump}.

At the same time, the rise of the anti-establishment right repelled a reactionary opposition towards the Democratic party \cite{hopkins2022trump}.  
This, to some degree, contributed to an increase in left pro-establishment publicly identifying users by 9\% (67,686 to 73,984) and left anti-establishment publicly identifying users by 6\% (64,367 to 68,075).  
The largest increases were ``Liberal'' by 23\% (34,793 to 42,968), ``Democrat'' by 36\% (11,762 to 16,052), and ``Progressive'' by 12\% (29,579 to 32,999). 
%(Note we do not have the data required to prove or disprove that these identities were added in direct reaction to the increase in ``MAGA'' and ``Trump'' identities).
 
The remaining political identities considered were relatively stable in their use in self-presentations or decreased immediately after the election.  Unsurprisingly, those that decreased after the election were directly related to candidate support (e.g., ``trump2016'', ``neverhillary'', ``nevertrump'').

Overall, we confirm previous findings that X users increasingly defined themselves politically between September 2016 and December 2018 \cite{rogers2021using}.  
However, this shift did not occur symmetrically across partisan and ideological groups.  
Rather, we see a substantial increase in public identification with anti-establishment right identities, accompanied by modest growth in left and pro-establishment right identities.    
%This increase was both due to political users joining X and existing users adding political identities.  

\subsubsection{Comparison of activity and influence}
While there was an increase in politically self-defined users, the proportion of users with any political identity in their profile in the 50 million user sample never exceeds 0.6\% between September 2016 and December 2018.
Given this statistic, one may conclude that this set of users is too small to make a significant impact.
To get a sense of their influence on X, we use unpaired t-tests to compare the average activity (tweet count) and following (follower count) of users that do \textit{not} have a political identifier in their bio with the following categories of users: right pro-establishment, right anti-establishment, left pro-establishment, left anti-establishment; see Table \ref{tab:follower_counts}, Table \ref{tab:tweet_counts}, and Table \ref{tab:t_tests} in the Appendix.  

In both September 2016 and December 2018, the users in all four categories of political identities tweet significantly ($p < 0.05$) more than users without any political identity in their bio.  Right pro-establishment, right anti-establishment, and left anti-establishment users have significantly ($p < 0.05$) more followers than users with no political identifiers, while left pro-establishment users do not ($p \geq 0.05$).
While self-defined pro-establishment left users have the most followers on average, they also vary the most.

%Figure \ref{fig2} in the Appendix contains a visual representation of the (log) mean, median, and standard error of follower count and tweet count of users in each of the five categories.  Left pro-establishment users have the most followers on average and variation in number of followers in both September 2016 (mean: 2778.17, std. error: 1373.13, median: 139.0) and December 2018 (mean: 3512.2, std. error: 1595.74, median: 159.0).  
%They also have the most tweets on average in September 2016 (mean: 6441.47, std. error: 85.23, median: 885.0).  
%In contrast, in December 2018, right anti-establishment users have more tweets on average than any other category of users (mean: 13,406.89, std. error: 112.66, median: 2,896.0).  
%By this time, there has been a consistent increase in ``MAGA'' supporters on X.  
%Trump was highly active on X and elicited more support (in this way) from his followers on the platform than other politicians \cite{kreis2017tweet}.  

In sum, the relatively small set of social media users with overt political signals have a disproportionate influence due to their high activity level (tweet count) and large influence (follower count) relative to non-political users.  We use this result as support for the idea that users could meaningfully affect the perceptions of alignment between identifiers and political identities of fellow X users.
%\footnote{We do not have the data required to assess actual exposure to these users.}

%\cite{barbera2015understanding,mukerjee2022political}  

%\subsection{Alignment of individual and types of identifiers}

%We now analyze the alignment of an assortment of individual and types of identifiers with the aim of providing a range of expected and surprising examples of social sorting in X profiles.

%We supplement findings at the individual identifier level with category level results to provide a more comprehensive picture of alignment with political identities on varying scales.

\subsection{Bias and variance of alignments over time}
\begin{figure*}[h!]
\centering
\includegraphics[width=\textwidth]{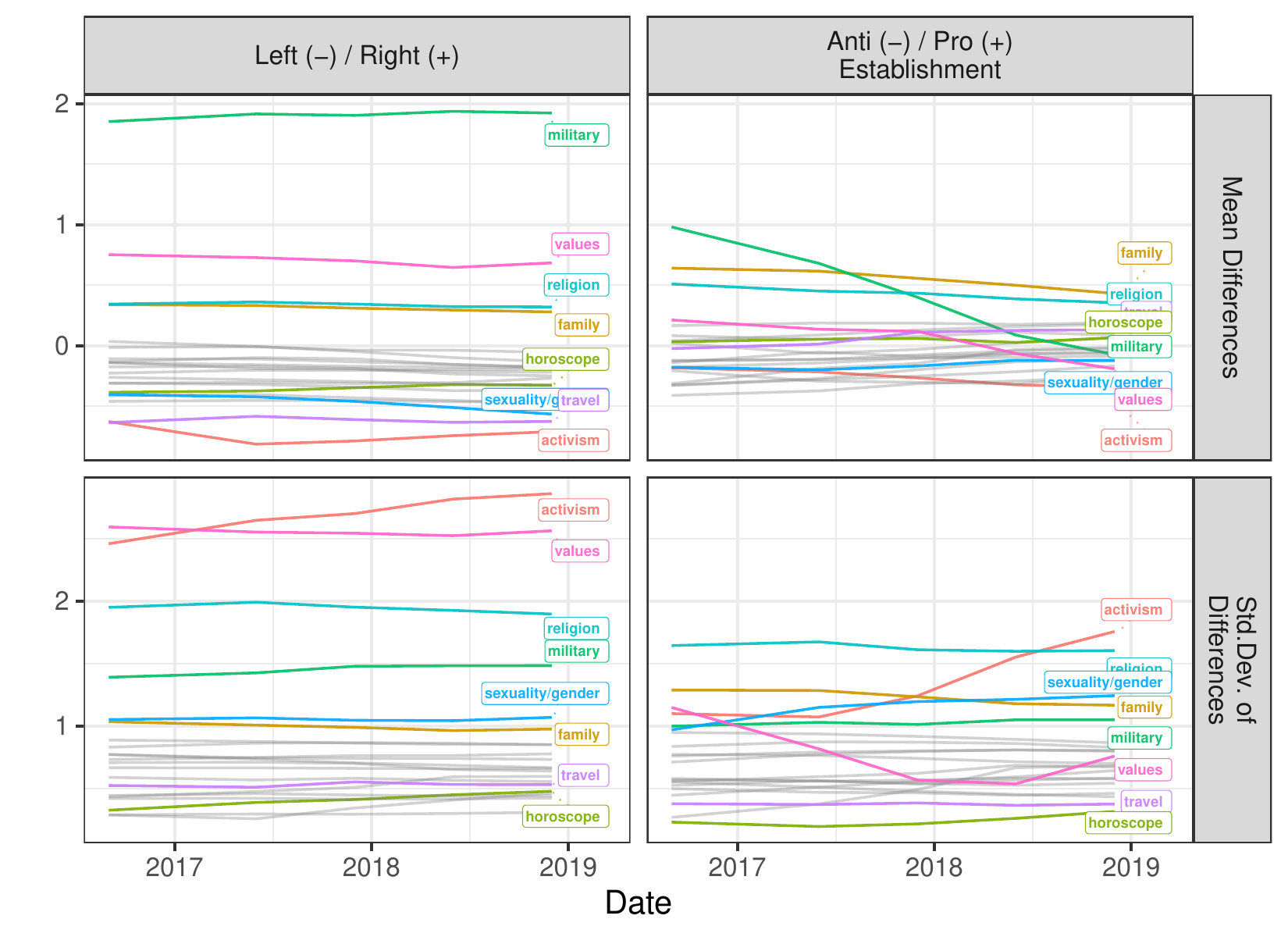}
\caption{Mean (top row) and standard deviation (bottom row) of differences in weighted log-odds between Right and Left (left column) and Pro- and Anti-establishment (right column).  Highlighted types of identifiers correspond to discussed results.}.
\label{fig34}
\end{figure*}

Figure \ref{fig34} represents the mean (left column) and standard deviation (right column) of z-scored differences in weighted log-odds between right and left (top row) and pro- and anti-establishment (bottom row) for each category of identifiers from September 2016 to December 2018; see Table G10 %\ref{tab:lr_sum} 
and Table G11
%\ref{tab:est_sum} 
in the Appendix for average and standard deviation for each category/time period.
The z-scored values incorporate the certainty we have in the point estimates by accounting for variance \cite{monroe2008fightin}.

Although most types of identifiers are not strongly biased towards either side of either dimension, there are a few notable exceptions.
The alignments of military, values, religion, and family identifiers are skewed towards the right on average, while disability/(mental) illness, culture, sexuality/gender, travel, and activism identifiers are skewed towards the left.
Along the pro-/anti-establishment dimension, family, religion, outdoors/nature/animals, and school/subject identifiers are skewed towards pro-establishment.  On the other hand, culture, activism, sex, and technology/gaming/cars identifiers are biased towards anti-establishment.

The identifiers that have the most variation in alignment between left and right identities are activism, values, religion and military.  
Between pro- and anti-establishment, alignment varies the most for activism (note the temporal increase in variation), religion, sexuality/gender, and family identifiers.  
These types of identifiers can be interpreted as the most aligned with one side or the other (i.e., ``polarized'').  

Notably, for military identifiers, the standard deviation of differences is smaller than the absolute value of the mean across time (mean: 1.90, std. dev.: 1.45).  This indicates that a majority of political users identifying with the military are on the right despite the relatively large standard deviation of differences.  
It is not that users on the left use different military identifiers than the right; users on the left largely do not use military identifiers at all.
The same holds for travel (mean: -0.62, std. dev.: 0.53) identifiers across time periods, except they are used largely by users with left-leaning identities and rarely by those with right-leaning ones.

Conversely, the alignment along the left/right dimension varies the least for age, relationship status, sports, horoscope, technology/gaming/cars, and sex identifiers.
Alignment along the pro-/anti-establishment dimension varies the least for horoscope, travel, sports, technology/gaming/cars, and sex.
This suggests certain identifiers used by politically self-defined users could decrease perceived differences between groups in some cases.

In terms of temporal shifts, the mean difference and standard deviation of differences are relatively stable between left and right users.  
There is much more movement in the values between pro- and anti-establishment, as the right anti-establishment emerged as distinct from the right pro-establishment.
In particular, the bias of military identifiers shifted from pro-establishment to anti-establishment between September 2016 and December 2018.
This likely indicates that users adding anti-establishment right identifiers to their profiles (i.e., ``MAGA'', ``Trump'') also include references to military affiliations in their profile.

In addition, the variance of alignments of activism identifiers notably increased (1.098 to 1.74) along the pro-/anti-establishment dimension throughout 2017 and 2018.
That is, identifiers that reference activism (e.g., ``ally'', ``community activist'', ``conservationist'', ``freemason'') were increasingly used overwhelmingly by one side or the other (for both dimensions).  
There are fewer common social causes people across political lines are willing to support and/or common ways people in different political groups represent their advocacy/support for advocacy.

On the other hand, the variance of differences in values between pro- and anti-establishment orientations decreased (1.18 to 0.78), although values are consistently highly divided along the left/right dimension.  
This is due, in part, to values that were included largely by pro-establishment right users (e.g., ``patriotic'', ``freedom'') being used more by the growing set of anti-establishment right users.

\subsection{Alignments of individual identifiers}
For individual identifiers, we report the difference in weighted log-odds with informative Dirichlet priors for each dimension, denoted ``LR'' and ``EST''.
We also report the probability $p$ of the observed (or more extreme) difference based on the distribution of differences to provide a sense of the strength of the alignment relative to other identifiers; where $^*: p<0.05$, $^{**}: p < 0.001$.
For clarity, we include the values from the sample taken in December 2017 unless otherwise specified.

Among the difference in weighted log-odds of the 3,838 selected identifiers, 254 (6.61\%) were statistically significant at the 0.05 level for the left/right dimension and 216 (5.62\%) pro-/anti-establishment dimension (115 identifiers were significant across both dimensions) in December 2017.  In total, 355 identities (9.2\%) were significant along at least one dimension. The number of statistically significant identifiers does not vary substantially between time periods.

Of course, ``statistical significance'' does not necessarily translate to theoretical significance.
However, we can conclude that most identifiers are weakly aligned with left/right and/or pro-/anti-establishment political identities, which means they are likely not strong contributors to subjective social sorting.
Moreover, we let theoretical interests drive the discussed results.

\paragraph{Religion}
Consistent with accounts of the religious right, ``Christian’’ (LR: 64.39$^{**}$; EST: $40.85^{**})$, ``God’’ (LR: 43.34$^{**}$; EST: $-2.48$), ``Jesus’’ (LR: 22.69$^{**}$; EST: $4.19$), and ``Catholic’’ (LR: 20.16$^{**}$; EST: $d=20.16^{**}$) are among the most strongly aligned identifiers with right partisan and ideological identities.  
In contrast, left identities are most aligned with non-traditional and non-religious identifiers, such as ``atheist’’ (LR: -35.38$^{**}$; EST: $4.64$, ``secular’’ (LR: 20.74$^{**}$; EST: $3.33$), ``spirituality’’ (LR: -11.98$^{*}$; EST: $6.47^{*}$), and ``agnostic'' (LR: -10.57$^{*}$; EST: $1.51$). 
In fact, most religions, excluding Christianity, are used (at least slightly) more in left-leaning profiles: ``buddhist'' (LR: -9.081; EST: -1.48), ``pagan'' (LR: -7.68; EST: -0.29), and ``muslim'' (LR: -7.55; EST: 3.0).  

%Along the pro-/anti-establishment dimension, alignments of religious identifiers are skewed pro-establishment overall, as shown in Figure \ref{fig34}.  
%However, there is a robust anti-establishment use of religious identifiers (e.g., ``pray'', ``islam'', ``pastor'', ``visionary'') causing the large spread in alignment.  
%There was not a pattern we could discern of religious identifiers used by one side or the other of this dimension; see ???.

\paragraph{Occupations/industry; military}
Left anti-establishment identifying users associate themselves the most with entertainment-related occupations, such as ``dj'' (LR: -21.53$^{**}$; EST: $-21.58^{**}$) and ``producer'' (LR: $-20.27^{**}$; EST: $-18.37^{**}$).  
Moreover, occupations in the education field are highly aligned with left-leaning identities overall, such as ``teacher'' (LR: $-14.44^{*}$; EST: $d = 3.72$) and ``educator'' (LR: $-13.81^{*}$; EST: $0.14$).
Professional, white collar, positions like ``lawyer'' (LR: $-11.24^{*}$; EST: $5.33^{*}$), ``scientist'' (LR: $-10.56^{*}$; EST: $-0.35$), and ``writer'' (LR: $-25.18^{**}$; EST: $-1.48$) are most aligned with left identities in X profiles as well.

On the other hand, users identifying on the right politically signal affiliations with the military and police, like ``military'' (LR: $28.66^{**}$; EST: $-2.84$), ``veteran'' (LR: 21.99$^{**}$; EST: $7.23^{*}$), and ``police'' (LR: 14.02$^{*}$; EST: $-3.16$).
Profiles of pro-establishment and anti-establishment users on the right are distinguished by the pro-establishment users' references to business leadership, such as ``small business owner'' (LR: 11.14$^{*}$; EST: 5.46$^{*}$) and ``chairman'' (LR: 11.66$^{*}$; EST: 12.6$^{**}$).

\paragraph{Activism}
Identifiers related to activism tend to be strongly aligned with one side of the left/right dimension and (to a lesser degree) one side of the pro-/anti-establishment dimension.  That is, both sides reference a distinct set of social and political movements.
For instance, references to specific Constitutional amendments, 1A (LR: 20.82$^{**}$; EST: -11.43$^{**}$) and 2A (LR: 42.62$^{**}$; EST: $-14.022^{**}$), signal support for and prioritization of free speech and gun rights by the anti-establishment right.

Some activism signals shared by users holding right-leaning political identities have a direct counterpart shared by left-leaning identity holders \cite{luders2022becoming}.  
A prime example of this is ``blacklivesmatter'' (LR: -22.72$^{**}$; EST: 0.16) vs. ``alllivesmatter'' (LR: 9.71$^{*}$; EST: -5.36$^{*}$) and ``bluelivesmatter'' (LR: 18.11$^{**}$; EST: -11.25$^{**}$) (again reflecting self-defined right-leaning users identification with and support for the police).
Self-presentation allows users to clearly indicate a stance on social movements, including countering stances they do not hold.

Other activism identifiers distinctly aligned with left identities include ``feminist'' (LR: -49.22$^{**}$; EST: 6.53$^{**}$), ``environmentalist'' (LR: $-16.69^*$; EST: $-1.679$), and ``resistance'' (LR: 16.52$^*$; EST:  0.79), which directly refers to supporting opposition to Trump\footnote{\url{https://www.france24.com/en/20180530-who-fighting-trump-opposition-meet-resistance-resist-twitter-hashtag-grassroots-usa}}. 
Trump affected the self-presentation of his greatest supporters \textit{and} haters on social media.

\paragraph{Family}
Interestingly, female family roles are highly aligned with pro-establishment identities across the left/right spectrum, like ``mom'' (LR: 1.15; EST: 15.55$^{**}$), ``grandmother'' (LR: 6.63; EST: 8.80$^{*}$), and ``sister''  (LR: -6.74; EST: 6.17$^{*}$).
Moreover, users that include right-leaning political identities in their profile prioritize expressing their affinity for their family by including identities like ``my family'' (LR: 11.79$^{*}$; EST: 0.087) and ``happily married'' (LR: 10.59$^{*}$; EST: 0.93), as well as family roles like ``father'' (LR: 4.70$^{**}$; EST: 16.74$^{**}$) and ``husband'' (LR: 20.92$^{**}$ ; EST: 16.88$^{**}$).

\paragraph{Gender/sexuality}
As expected, users with left-leaning identities in their profile are much more likely to reveal their sexual orientation and gender preferences in their self-presentation.  Commonly used sexuality signals are ``lgbt'' (LR: $-17.8^{**}$; EST: 2.94) and ``gay'' (LR: $-16.51^{**}$; EST: 1.45).
The most left aligned gender preferences are ``she'' (LR: $-11.71^{*}$; EST: 0.83) and ``her'' (LR: -8.36$^{*}$; EST: 2.36).

\paragraph{Nationality/locations}
Although nationalities and locations are not highly sorted overall, there is a small set of them that differentiate left and right identity holders.  
On the left, the most highly aligned identifiers in this category include ``indian'' (LR: -11.15$^{*}$; EST: 2.096), ``pakistan'' (LR: $10.99^{**}$; EST: $-1.5$), and ``european'' (LR: $-10.6^{*}$; EST: $1.77$). 
On the right, highly aligned identifiers include ``american'' (LR: 35.35$^{**}$; EST: 7.015$^{*}$), ``israel'' (LR: 25.86$^{**}$; EST: 3.086), and ``texan'' (LR: 12.54$^{*}$; EST: 9.67$^{*}$).

\paragraph{Values}
Values expressed in X profiles are much more divided by the left/right dimension than pro-/anti-establishment, reflecting the dominant role of traditional partisanship.
Values like ``patriot'' (LR: $39.38^{**}$; EST: $3.66$), ``freedom'' (LR: $22.30^{**}$; EST: $2.02$), and ``liberty'' (LR: 20.41$^{**}$; EST: $3.06$) are highly aligned with right-oriented identities.
On the other hand, users with left-oriented identities in their profile are more likely to also include value identifiers like ``peace'' (LR: $-11.67^{*}$; EST: $-3.09$) and ``humanity'' (LR: $-7.45$; EST: 0.11).

\paragraph{Platform afforded and event driven}
Some highly aligned identifiers reference specific events or platform affordances. 
For example, ``deplorable'' was used by Hillary Clinton to describe Trump supporters in 2016, which they ultimately embraced\footnote{\url{https://www.usatoday.com/story/news/politics/onpolitics/2016/09/12/deplorable-and-proud-some-trump-supporters-embrace-label/90290760/}}.  
This resulted in both on and offline inclusion of ``deplorable'' in the self-presentation of holders of anti-establishment right identities (LR: 36.47$^{**}$; EST: -24.79$^{**}$).
Anti-establishment right identities and ``covfefe'' (LR: 13.32$^{*}$; EST: -12.16$^{**}$), which references a typo in one of Trump's tweets, are also highly aligned.

Another example is the strong alignment between pro-establishment right identities and ``tgdn'' (LR: 12.32$^{*}$; EST: 9.04$^{*}$), which refers to the Twitter Gulag Defense Force\footnote{\url{https://www.socialmediatoday.com/content/social-advocacy-and-politics-twitter-failing-policing-troll-traps}}.
The TGDN, which as since largely disappeared from social media, sought to troll users expressing left-leaning ideas and identities.
These profiles of these accounts may or may not represent the person behind the account, perhaps most likely not if the goal is to troll other users, but still impact how other users perceive the pro-establishment right.
In this case, they are associated with the behaviors of users acting in the name of the TGDN.

\paragraph{Weak alignment}
Equally as important as the strongest alignments, but not the main focus of this paper, are the weakest alignments between identifiers and political identities.
Such identifiers represent commonalities between even the most extreme partisans.
We provide a brief assessment of these identifiers here.

Commonly used values across both dimensions include ``independence'' (LR:  $-0.19$; EST: $-0.49$) and ``hedonist'' (LR: $-1.82$; EST: $-0.62$).
There are also activism identifiers that bridge political groups, like ``world changer'' (LR: -0.11; EST: 0.23), ``community volunteer'' (LR: -0.83; EST: 1.81), and ``conservationist'' (LR: -1.034; EST: 1.43).

Unsurprisingly, affinities for nature, animals, and sports also traverse political divides, such as ``loves nature'' (LR: -0.002; EST: -1.33), ``loves cats'' (LR: -0.026; EST: 0.44), ``scuba diving'' (LR: -0.33; EST: -0.16), and ``baseball'' (LR: -0.087; EST: 1.59).
More surprising are the (social) media affiliations and preferences included in the profiles of users across political groups, like ``cnn'' (LR: 4.1; EST: -3.84), ``nytimes'' (LR: -0.3; EST: -1.11), and ``columbiajourn'' (LR: -0.17; EST: -0.047).

\section{Discussion}\label{sec5}

In this work, we investigated if and how self-presentations afforded by social media profiles may exasperate subjective social sorting along political lines.  
To do so, we analyzed the alignment between identifiers (social identities, preferences, affiliations) and political identities defined by two relevant dimensions---left/right and pro-/anti-establishment orientations---in X user bios.
The answer, as one may expect, is: it depends.

We found the majority (approx. 90\%) of identifiers do not strongly align along either dimension in user bios.
The remaining identifiers significantly align along at least one dimension, dominantly along the left/right dimension, indicating there are persistent patterns of co-occurrence between certain identifiers and political identities in self-presentations on social media.
That is, subjective social sorting of identifiers along political lines is reflected to some degree in X user profiles.
Many of these strong alignments reinforce existing associations (e.g., ``lgbt'' and left, ``Christian'' and right), while others reflect social media afforded activism (e.g., ``deplorable'' and anti-establishment right, ``TGDN'' and pro-establishment right).  

%Strong alignments can enhance polarization because it validates misperceptions that, say, all people on the right are fathers or all people on the left are vegan, thus increasing perceived division and enhancing outgroup hostility \cite{druckman2022mis}.  
%This could, in part, drive tendencies to overestimate the proportion of partisan group members that belong to party-stereotypical groups \cite{ahler2018parties,druckman2022mis}.

We deepened our investigation by categorizing identifiers into 21 non-political categories (\textit{RQ2a}) and examining the evolution of alignments between September 2016 (pre-Trump presidency) and December 2018 (mid-Trump presidency) (\textit{RQ2b}).

Some types of identifiers may be especially effective for bridging algorithms and other depolarization efforts to center around due to their lack of variance across groups: sports, technology/gaming/cars, travel, age, sex, and relationship status.  
When attempting to reduce outgroup hostility, emphasizing common identities and/or activities in these domains may be more likely to have positive results than, say, religion.
Indeed, research shows depolarization interventions that emphasize common identities across partisan groups are most effective in decreasing partisan animosity (i.e., affective polarization) \cite{voelkel2023megastudy}.

We also found types of non-political identifiers that emphasize differences between groups due to strong alignment with one side or the other: religion, activism, values, family, military, and sexuality/gender.
This is especially concerning given there are appreciably more users with non-political identifiers in their profiles relative to political ones (e.g., family identifiers are in more than 4M user profiles of the 50M sample vs. $\sim$300,000 users with political identities); see Table E4 %\ref{tab:category_user_counts} 
and E5
%\ref{tab:category_user_avg} 
in the Appendix.
It follows that most identifiers are largely used in profiles that do \textit{not} contain a political identity.  
Attitudes and behaviors directed towards users with profiles that only contain non-political identifiers, if they are perceived as highly aligned with certain political identities, can be affected by pre-existing notions of that political group \cite{brewer2005social,roccas2002social}.
If the associated political group is opposed to the viewing user's position, we expect this results in increased toxicity and hostility between social media users due to affective polarization.

Crucially, alignments in X user profiles can only affect perceptions of social sorting and polarization if other users actually see them.
In addressing \textit{RQ1}, we confirmed recent results that X users increasingly defined themselves politically throughout this time period \cite{rogers2021using}.
In addition, we showed self-defined political users tend to be more active and popular on X than the average user.
However, the true influence of self-defined political users on X is unclear, as the vast majority of users do not directly engage with American politics on the platform \cite{mukerjee2022political}.
Without access to additional data (e.g., exposure to and dwell time on political user profiles), it is challenging to estimate the influence of self-presentations on lasting behavioral and attitudinal changes.
It may be that social sorting observed in self-defined political user profiles mostly influences other strong partisans/ideologues.

We also demonstrated temporal shifts in partisan alignment due to offline events.
Donald Trump played a pivotal role in the rise in populist rhetoric and attitudes in 2016 \cite{oliver2016rise,hopkins2022trump}.
He served as a catalyst for party change, drawing people who hold anti-establishment and anti-elite attitudes towards the Republican party and repelling reactionary opposition towards the Democratic party \cite{hopkins2022trump}.
The increase in co-occurrences of pro- and anti-establishment right identities in profiles throughout this time period shows users adding ``MAGA'' and ``Trump'' tend to emerge from and/or join the Republican party.
In addition, the unprecedented growth in anti-establishment right identities in X user profiles affected alignments, as certain types of identifiers (i.e., family, religion, and military) that were biased strongly towards the pro-establishment right in September 2016 shifted towards the anti-establishment right throughout 2017 and 2018.

%The social sorting observed in the profiles of these users reflects and possibly extends the signals from elected officials, mainstream media, and political pundits about the norms of that partisan and ideological group. 

Critically, the alignments discussed in this work are \textbf{not} representative of objective social sorting, and should not be interpreted as such.
Users with political identities in their profile are \textbf{not} representative of all members of that political group.
For example, they tend to have stronger ties to political groups and ideas than users who do not public identify politically \cite{rainie2012social}.
%Our results do not imply that there are more differences between political groups than previously known in the general public.

Collectively, these results indicate social sorting is reflected in X user profiles to some degree.  
%We systematically identified personal identifiers in a 50M user sample and quantified political alignment of identifiers (social identities, preferences, affiliations) along left/right and pro-/anti-establishment dimensions.
%In this way, we can identify which individual and types of identifiers are the most and least aligned along partisan lines and over time.
This work advances theory by proposing and demonstrating a pathway for self-presentation on social media to facilitate detrimental social sorting along partisan identities.
In addition, the results directly inform practical interventions for partisan animosity by suggesting effective topics to use when highlighting commonalities between groups in bridging interventions \cite{voelkel2023megastudy} and algorithms \cite{ovadya2022bridging}.

%Social identities can be politicized \cite{mason2015disrespectfully,mason2016cross} just like subreddits \cite{waller2021quantifying} , beliefs, phrases \cite{monroe2008fightin}, and ideas.  
%Although partisan alignment of social identities, preferences, and affiliations do not directly make people more extreme ideologically, it affects information exposure and sharing due to changes in social networks, as well as the influence of cognitive biases due to shifts in ingroup and outgroup attitudes.

\subsection{Limitations}
In order to see changes in the social meaning of individual identifiers, we likely need to consider longer time horizons than two years.  For example, studies of partisan alignment of beliefs stretch decades \cite{dellaposta2020pluralistic}.  

In addition, there are always opportunities for error when relying on human annotations, as we did to select identifiers from phrases in bios and categorize identifiers into 21 categories.
We hope other researchers will use and refine these labels in their work.
Relatedly, our assignments of partisan and ideological identities into categories defined by left/right and pro-/anti-establishment orientations required subjective assessments that some scholars may disagree with. 
%In full transparency, we are the least confident in the identities designated as left and anti-establishment.
We provide the data and code required to run the same analyses with different combinations of political identities (e.g., separate partisan and ideological identities, remove politicians).

We do not attempt to distinguish between bots and human users because we are characterizing the profiles that are shaping perceptions of members of political groups.
Humans have limited ability to discern (undisclosed) bots from human run accounts \cite{kenny2024duped}, so we do not expect bot-like behavior (recognized by automated tools, e.g., \cite{ng2023botbuster}) to substantially affect interpretation of co-occurrences of identifiers in profiles.

%We may have missed niche identifiers since we only considered those used by more than 1000 users.  We also did not consider emojis, which are certainly markers of identity on social media \cite{li2020emoji}.

More research is needed to understand the short term and long term cognitive and social impacts of social sorting on social media.  Do users actually start to avoid/seek users holding non-political identifiers they perceive as being associated with a certain political identity?  
How much exposure (and in what context) to co-occurring social and political identities is required for users to consistently associate them together?
%In other words, does subjective social sorting facilitated by alignments between personal identities and political identities in social media profiles actually facilitate polarizing behavioral and attitudinal shifts?

\backmatter

%\bmhead{Supplementary information}

%\bmhead{Acknowledgements}

%\section*{Declarations}

%Some journals require declarations to be submitted in a standardised format. Please check the Instructions for Authors of the journal to which you are submitting to see if you need to complete this section. If yes, your manuscript must contain the following sections under the heading `Declarations':

%\begin{itemize}
%\item Conflict of interest/Competing interests (check journal-specific guidelines for which heading to use)
%\item Ethics approval and consent to participate
%\item Consent for publication
%\item Data availability 
%\end{itemize}

%\noindent
%If any of the sections are not relevant to your manuscript, please include the heading and write `Not applicable' for that section. 


%Springer journals and proceedings: \url{https://www.springer.com/gp/editorial-policies}

\begin{appendices}

\section{FAIR Compliance}
In compliance with FAIR principles, the original data used in this work is \textit{findable} upon request to the corresponding author (more information, including unidentifiable derivatives of the data and relevant Datasheets, are available via OSF\footnote{OSF LINK to be added}), \textit{accessible} because it is in standard data formats (.csv, .json), \textit{interoperable} because nearly any programming language has libraries that can be used to access the datasets in these standard formats, and \textit{reusable} due to the richness of the dataset itself and in-depth descriptive information provided in this paper and the Datasheet on the OSF repository.

\section{Co-occurrence network summary statistics}

Table \ref{tab:co_occur} contains the summary statistics for the co-occurrence networks between identifiers generated at each time period.

\begin{table}[]
    \centering
    \begin{tabular}{lccccc}
    \toprule
& September 2016 & June 2017 & December 2017 & June 2018 & December 2018 \\
\midrule
Number of nodes & 3668 & 3670 &  3670 & 3670 & 3670 \\
\midrule 
Number of edges & 4061169 & 4070042 & 4071017 & 4063726 & 4057417 \\
\midrule
Density & 0.302 & 0.302 & 0.302 & 0.302 & 0.301 \\
\midrule
Average (std. dev.) of weights & 43.26 (462.24) & 42.99 (455.97) & 43.02 (454.79) & 42.93 (452.83) & 42.98 (453.59) \\
\midrule
Median weight & 4 & 4 & 4 & 4 &  4\\
\bottomrule
    \end{tabular}
    \caption{Summary statistics for co-occurrence networks for each time period.}
    \label{tab:co_occur}
\end{table}

\section{Personal identifier classifier}

Our personal identifier classifier was developed in two steps. First, we embedded all phrases extracted from bios into a 1024-dimensional embedding space using model from \citet{madani2023measuring}, who fine-tune a Sentence-BERT model using a large sample of Twitter bios. Their work shows that this embedding produces better alignment with human perception on a number of evaluation tasks.
%; here, we find that our classification accuracy increases on the order of 10-20\% (depending on the model, see below) by using their embeddings over a non-fine-tuned embedding model. 

Given these embeddings, we then train a model to predict whether or not a given phrase extracted by by the method from \citet{pathak2021method} is a personal identifier or not. To do so, we use the 2,271 manually labeled samples as training data, with the manual labels of personal identifier or not as the outcome variable and the embeddings from Madani et al.'s (2023) method as the features. We train a number of different standard machine learning models with this setup using 10-fold cross validation. 
Selecting the best model from these initial runs (SVM), we evaluate the performance for all combinations of following hyperparameter values: $C$=[0.1, 1, 10, 100, 1000], gamma = [1, 0.1, 0.01, 0.001, 0.0001], kernal = ['linear', 'poly', 'rbf', 'sigmoid'].
The model performs the best for $C=10$, gamma=1, kernal = 'rbf'.
The accuracy, precision, and macro-F1 score for each model is in Table \ref{tab:ml_stats}.
This model was used in this work and will be shared upon acceptance.
Model training took under ten minutes on a personal laptop (Apple M1 chip).

%Potential consequences of misclassification include inclusion of phrases that are not personal identifiers in aggregate alignment measures, which could lead to incorrect interpretations of alignments in X profiles.

\begin{table}[h]
    \centering
    \begin{tabular}{lcccc}
\toprule
& \multicolumn{4}{c}{Fine-tuned model} \\
& Accuracy & Precision & Macro-F1 & Recall \\
\midrule
SVM  (without hyperparameter tuning) &   0.857 (0.0072) & 0.844 (0.014) & 0.856 (0.0072) &  0.888 (0.011) \\
\midrule
SVM  (with hyperparameter tuning) &   \textbf{0.878} (0.0066) & \textbf{0.868} (0.014) & \textbf{0.878} (0.0064) &  \textbf{0.903} (0.016)) \\
\midrule
Logistic regression    & 0.827 (0.011) & 0.833 (0.014) & 0.826 (0.011) &  0.841 (0.025) \\
\midrule
Decision tree & 0.791 (0.011) & 0.799 (0.015) & 0.791 (0.011) &  0.793 (0.017)   \\
\midrule
Naive Bayes  &  0.784 (0.0049) & 0.762 (0.016) & 0.783 (0.0087) & 0.832 (0.0089)  \\
\midrule
K-nearest neighbors (k=3)  &  0.85 (0.0069) & 0.845 (0.014) & 0.849 (0.007) &  0.872 (0.012)  \\
\midrule
K-nearest neighbors (k=5) &   0.849 (0.0091) & 0.833 (0.012) & 0.848 (0.0093) &  0.886 (0.01) \\
\midrule
K-nearest neighbors (k=10) &   0.845 (0.012) & 0.842 (0.0098) & 0.845 (0.012) &  0.861 (0.025)  \\
\midrule
K-nearest neighbors (k=15)  & 0.840 (0.0092) & 0.822 (0.017) & 0.839 (0.0094) &  0.886 (0.015)  \\
\bottomrule
    \end{tabular}
    \caption{Average (standard deviation) of accuracy, precision, macro-F1, and recall score for a set of classical machine learning models.}
    \label{tab:ml_stats}
\end{table}

\section{Identifier category annotation statistics.}

Table 4
%\ref{tab:category_counts}
contains the number of identifiers assigned to each category, as well as the Krippendorf's alpha for the subset of identifiers labelled by two annotators.  

\begin{table}[t]
\begin{tabular}{@{}p{0.55\linewidth} | p{0.14\linewidth}| p{0.27\linewidth}}
\hline
Category &  Number labelled & Krippendorf's Alpha (number labelled by two annotators) \\
\hline
political & 30 & NA \\
\hline
culture & 646 & 0.793 (94) \\
\hline
occupation/industry & 444 & 0.81 (63) \\
\hline
sports  & 425 & 0.834 (59) \\
\hline
nationality/location/ethnicity/race & 377 & 0.935 (46)\\
\hline
school/subjects & 231 & 0.732 (35) \\
\hline
technology/gaming/cars & 171 & 0.812 (23) \\
\hline
outdoors/nature/animals  & 129 & 0.755 (21) \\
\hline
family & 105 & 0.940 (9)\\
\hline
religion & 104 & 1.0 (9)\\
\hline
business/finance & 66 & 0.907 (6) \\
\hline
activism & 63 & 0.765 (8) \\
\hline
(social) media & 62 & 0.841 (15)\\
\hline
sexuality/gender & 61 & 0.665 (2) \\
\hline
age & 42 & 0.797 (9) \\
\hline
disability/(mental) illness &  31 & 1.0 (5) \\
\hline
horoscope & 26 & 1.0 (3) \\
\hline
travel & 21 & 1.0 (4) \\
\hline
sex & 20 & 0.799 (3) \\
\hline
military & 17 & 0.856 (4) \\
\hline
relationship status & 17 & 0.799 (3) \\
\hline
values & 17 & 1.0 (1) \\
\hline 
\end{tabular}
\label{tab:category_counts}
\caption{Number of identifiers labelled as each category.}
\end{table}

\section{Number of users with each type of identifier in their profile in each time period.}

Table E4 %\ref{tab:category_user_counts} 
contains the \emph{total} number of users who use at least one identifier for each category and time period.
Because the number of identifiers assigned to each category varies between 17 and 646, we also report the \emph{average number of users per identifier}  for each category in Table
E5.
%\ref{tab:category_user_avg}.

\begin{table}[h]
\begin{tabular}{@{}lccccc}
\hline
Category &  September 2016 & June 2017 & December 2017 & June 2018 & December 2018 \\
\hline
political & 272,009 & 293,320 & 301,854 & 309,263 & 320,879 \\
\hline 
sports & 5,188,787 & 5,074,286 & 5,029,249 & 4,963,088 & 4,910,120 \\
\hline 
occupation/industry & 8,069,514 & 8,135,294 & 8,200,694 & 8,204,900 & 8,240,269 \\
\hline 
school/subjects & 5,040,143 & 5,008,702 & 5,022,054 & 4,999,465 & 4,990,851 \\
\hline 
age & 1,567,282 & 1,490,253 & 1,463,448 & 1,422,964 & 1,396,297 \\
\hline 
culture & 10,933,925 & 10,778,951 & 10,721,508 & 10,624,169 & 10,547,797 \\
\hline 
technology/gaming/cars & 3,466,441 & 3,442,905 & 3,438,044 & 3,416,792 & 3,407,351 \\
\hline 
nationality/location/ethnicity/race & 7,183,715 & 7,155,498 & 7,159,897 & 7,122,230 & 7,097,553 \\
\hline 
family & 4,208,542 & 4,237,136 & 4,279,602 & 4,305,535 & 4,341,620 \\
\hline 
outdoors/nature/animals & 2,081,455 & 2,088,283 & 2,101,564 & 2,107,127 & 2,111,780 \\
\hline 
relationship status & 843,710 & 812,287 & 797,619 & 783,220 & 771,078 \\
\hline 
disability/(mental) illness & 350,504 & 352,726 & 355,017 & 355,767 & 357,859 \\
\hline 
religion & 2,865,652 & 2,840,731 & 2,832,161 & 2,808,222 & 2,788,663 \\
\hline 
sexuality/gender & 3,992,506 & 3,934,800 & 3,920,618 & 3,899,229 & 3,913,509 \\
\hline 
travel & 686,207 & 692,981 & 695,799 & 694,658 & 693,843 \\
\hline 
(social) media & 1,692,985 & 1,676,631 & 1,666,596 & 1,650,617 & 1,639,394 \\
\hline 
business/finance & 2,632,527 & 2,660,719 & 2,682,688 & 2,683,359 & 2,691,727 \\
\hline 
military & 332,125 & 343,397 & 352,355 & 358,058 & 362,792 \\
\hline 
activism & 729,378 & 767,422 & 782,429 & 792,972 & 799,442 \\
\hline 
horoscope & 375,562 & 371,637 & 372,929 & 375,997 & 380,986 \\
\hline 
sex & 279,677 & 275,385 & 274,468 & 273,412 & 273,715 \\
\hline 
values & 351,732 & 357,847 & 361,584 & 361,623 & 363,049 \\
\hline 
\end{tabular}
\label{tab:category_user_counts}
\caption{Number of users with a profile containing at least one identifier of each category.}
\end{table}

\begin{table}[!h]
\label{tab:category_user_avg}
\centering
\begin{tabular}{@{}lccccc}
\hline
Category &  September 2016 & June 2017 & December 2017 & June 2018 & December 2018 \\
\hline
political & 9066.97 & 9777.33 & 10061.8 & 10308.77 & 10695.96
\\
\hline 
sports & 12208.91 & 11939.50 & 11833.53 & 11677.85 & 11553.22 \\
\hline 
occupation/industry & 18174.58 & 18322.73 & 18470.032 & 8479.50 & 18559.16 \\
\hline 
school/subjects & 21818.80 & 21682.69 & 21740.49 & 21642.70 & 21605.41 \\
\hline 
age & 37316.24 & 35482.21 & 34844.00 & 33880.095 & 33245.17 \\
\hline 
culture & 16925.58 & 16685.68 & 16596.76 & 16446.082 & 16327.86 \\
\hline 
technology/gaming/cars & 20271.58 & 20133.95 & 20105.52 & 19981.24 & 19926.029 \\
\hline 
nationality/location/ethnicity/race & 19054.95 & 18980.10 & 18991.77 & 18891.86 & 18826.40 \\
\hline 
family & 40081.35 & 40353.68 & 40758.11 & 41005.095 & 41348.76 \\
\hline 
outdoors/nature/animals & 16135.31 & 16188.24 & 16291.19 & 16334.32 & 16370.39 \\
\hline 
relationship status & 49630.0 & 47781.59 & 46918.76 & 46071.76 & 45357.53 \\
\hline 
disability/(mental) illness & 11306.58 & 11378.26 & 11452.16 & 11476.35 & 11543.84 \\
\hline 
religion & 27554.35 & 27314.72 & 27232.32 & 27002.13 & 26814.067 \\
\hline 
sexuality/gender & 65450.92 & 64504.92 & 64272.43 & 63921.79 & 64155.88 \\
\hline 
travel & 32676.52 & 32999.095 & 33133.28 & 33078.95 & 33040.14 \\
\hline 
(social) media & 27306.21 & 27042.43 & 26880.58 & 26622.85 & 26441.84 \\
\hline 
business/finance & 39886.77 & 40313.92 & 40646.79 & 40656.95 & 40783.74 \\
\hline 
military & 19536.76 & 20199.82 & 20726.76 & 21062.23 & 21340.70 \\
\hline 
activism & 11577.43 & 12181.30 & 12419.51 & 12586.86 & 12689.55 \\
\hline 
horoscope & 14444.69 & 14293.73 & 14343.42 & 14461.42 & 14653.31 \\
\hline 
sex & 13983.85 & 13769.25 & 13723.4 & 13670.6 & 13685.75 \\
\hline 
values & 20690.12 & 21049.82 & 21269.65 & 21271.94 & 21355.82 \\
\hline 
\end{tabular}
\caption{Average number of users per identifier in each category.}
\end{table}

\section{Counts, follower counts, and tweet counts of users.}

Table \ref{tab:counts} contains the number of users with 1) each category of political identity, 2) any political identities, and 3) no political identities in their profile. 
Table \ref{tab:follower_counts} and Table \ref{tab:tweet_counts} contains the mean, standard deviation, and median number of followers and tweets, respectively, for users with each category of political identity, any, or no political identities in their profile.

In total, we ran sixteen unpaired Welch's t-tests with Bonferroni corrections--- four to compare tweet counts and four to compare follower counts of users with and without political identities in their profile in two different tiem periods (September 2016, December 2018).  
We use Welch's t-test because the standard deviations of tweet and follower counts, as well as the sample size, across groups vary substantially as shown in Tables \ref{tab:counts}, \ref{tab:follower_counts}, and \ref{tab:tweet_counts}.  The sample tweet count and follower count distributions are not normal (skewed right).  The other assumptions of Welch's t-tests are (at least approximately) fulfilled: continuous outcome, categorical explanatory variable, independent observations.

We use data collected in September 2016 and Demember 2018 for comparisons and do not observe major temporal shifts in differences between groups of users.  Corrected and not corrected p-values for each t-test can be found in Table \ref{tab:t_tests}.

We do not have a strong theoretical explanation for the demonstrated difference in activity level between political and non-political users.
The purpose of these hypothesis tests was to quickly evaluate \textit{if} self-defined political users differ from the rest of the users in our sample in tweet or follower count on average.

We know users with political identities in their profile are more likely to engage in civic and political discourse on X \cite{rainie2012social}.
It may be that these users have clearer goals for their use of the platform that lead them to tweet more often (e.g., encourage people to vote, solicit support for a candidate).
In contrast, the majority of X users are ``lurkers'', that is, they rarely actually tweet\footnote{\url{https://www.pewresearch.org/short-reads/2022/03/16/5-facts-about-twitter-lurkers/}}.
%Moreover, more tweets is associated with higher follower counts on average \cite{mueller2017predicting}.
More research is required to assess the underlying mechanism(s) behind observed differences in tweet/follower count and inclusion of political identities in profiles, which we do not address in this work.

%In sum, more research is required to assess the underlying mechanism(s) behind observed differences in tweet/follower count and inclusion of political identities in profiles.
%This could be done by surveying a sample of users who do and do not include political identities in their social media profile about why and when they tweet, as well as their overarching goals for using the platform.  
%In addition, investigating people's perceptions of accounts with a description that does and does not include political identities would shed light on the observed relationship between political identities in profiles and follower counts.

%\begin{figure*}[h!]
%\centering
%\includegraphics[width=2\columnwidth]{fig2.png}
%\caption{Mean (with standard error bars) and median follower and tweet counts for users in each political identity category in September 2016 and December 2018.}.
%\label{fig2}
%\end{figure*}

\begin{table}[]
    \centering
    \begin{tabular}{llllll}
\toprule 
Political identity type & September 2016 & June 2017 & December 2017 & June 2018 & December 2018 \\
\midrule
Right/pro & 80654 & 84141 & 85622 & 87873 & 90318 \\
\hline
Right/anti & 70129 & 71002 & 92746 & 100005 & 109696 \\
\hline
Left/pro & 67686 & 70807 & 72050 & 72600 & 73984 \\
\hline
Left/anti & 64367 & 66330 & 66849 & 67147 & 68075 \\ 
\hline
Any political identity & 272009 & 293320 & 301854 & 309263 & 320879 \\ 
\hline
English bio but no political identity & 50767589 & 50404055 & 50313603 & 49961251 & 49757784  \\
\bottomrule
\end{tabular}
 \caption{Number of users with each category of political identity in their profile in each time period.}
    \label{tab:counts}
\end{table}

\begin{table}[]
    \centering
\begin{tabular}{lp{2cm}p{2cm}p{2cm}p{2cm}p{2cm}}
\toprule
Political identity type & September 2016 & June 2017 & December 2017 & June 2018 & December 2018 \\
\midrule 
Right/pro & 1047.021, 10985.29, 153.0 & 1175.62, 20325.3, 158.0 & 1244.78, 12660.095, 163.0 & 1325.58, 14154.09, 170.0 & 1418.22, 12851.89, 175.0  \\ 
\hline
Right/anti & 1094.092, 14109.09, 137.0 & 2024.82, 122021.82, 201.0 & 2265.13, 148831.7, 194.0 & 2506.0, 168053.8, 227.0  & 2769.22, 171971.4, 270.0  \\
\hline
Left/pro & 2685.43, 307045.7, 135.0 & 3166.97, 373011.69, 144.0 & 3399.47, 381595.1, 144.0 & 3571.34, 400609.9, 150.0 & 3579.45, 399380.5, 151.0 \\
\hline
Left/anti & 932.2, 19574.01, 124.0 & 1014.83, 19137.48, 134.0 & 1107.40, 23342.30, 139.0 &  1204.76, 30932.39, 141.0 & 1112.55, 26244.95, 142.0  \\
\hline
Any political identity & 1428.90, 153732.45, 137.0 & 1794.86, 187302.87, 146.0 & 2011.14, 204269.86, 158.0 & 2157.69, 216952.013, 167.0 & 2232.015, 216955.7, 174.0 \\
\hline
English bio but no political identity & 724.49, 48158.32, 82.0 & 792.39, 51663.33, 86.0 & 846.011, 52840.21, 88.0 & 873.44, 56725.66, 88.0 & 838.51, 57726.20, 83.0  \\
\bottomrule
\end{tabular}
    \caption{Mean, median, standard deviation of follower count for users with each category of political identity in their profile in each time period.}
    \label{tab:follower_counts}
\end{table}

\begin{table}[]
    \centering
\begin{tabular}{lp{2cm}p{2cm}p{2cm}p{2cm}p{2cm}}
\toprule 
Political identity type & September 2016 & June 2017 & December 2017 & June 2018 & December 2018  \\
\midrule
Right/pro & 5374.074, 20173.52, 741.0 & 6299.64, 21963.63, 920.0 & 6971.82, 23976.12, 1030.0 & 7538.40, 25088.54, 1130.0 & 8378.4, 27488.92, 1263.0 \\
\hline
Right/anti & 5115.012, 18855.33, 626.0 & 7363.14, 23371.54, 1096.0 & 9095.823, 26871.28, 1470.0 & 10409.15, 29562.69, 1788.0 & 12125.54, 33399.02, 2244.0 \\
\hline
Left/pro & 6161.004, 19751.98, 810.0 & 6711.75, 21324.69, 1001.0 & 7590.33, 23972.12, 1156.0 & 8363.51, 25805.13, 1270.0  & 9096.025, 28745.10, 1391.0  \\
\hline
Left/anti & 3955.35, 14166.27, 451.0 & 4529.066, 15826.15, 558.0 & 5040.28, 17617.51, 623.0 & 5413.72, 18876.96, 669.00  & 5878.47, 20170.11, 736.0 \\
\hline
Any political identity & 5151.99, 18639.95, 640.0 & 6277.43, 21456.0, 861.0 & 7192.072, 23682.94, 1012.0 & 7924.94, 25400.22, 1130.0 & 8905.87, 28211.64, 1295.0 \\
\hline
English bio but no political identity & 2893.43, 10456.57, 250.0 & 3102.80, 11352.046, 286.0 & 3248.78, 11979.96, 305.0 & 3354.55, 12622.092, 316.0 & 3459.015, 12959.94, 327.0  \\
\bottomrule
\end{tabular}
    \caption{Mean, median, standard deviation tweet count for users with each category of political identity in their profile in each time period.}
    \label{tab:tweet_counts}
\end{table}

\begin{table}[]
    \centering
    \begin{tabular}{lccccc}
\toprule
    Date & Feature & Political identity type & T-statistic & Un-adjusted p-value & Adjusted p-value \\
 \hline   
 & Tweet count & Right/pro & 34.91 & $<$1e-20 & $<$1e-20 \\
%\cline{2-6}
 & Tweet count & Right/anti & 31.19 & $<$1e-20 &  $<$1e-20\\
%\cline{2-6}
& Tweet count & Left/pro & 43.031 & $<$1e-20 &  $<$1e-20\\
%\cline{2-6}
\multirow{8}{4em}{September 2016} & Tweet count & Left/anti & 19.012 & $<$1e-20 & $<$1e-20 \\
%\cline{2-6}
 & Follower count & Right/pro & 8.21 & 2.17e-16 & 8.70e-16 \\
%\cline{2-6}
 & Follower count & Right/anti & 6.88 & 5.95e-12 &  2.38e-11 \\
%\cline{2-6}
 & Follower count & Left/pro & 1.66 & 0.097 & 0.39  \\
%\cline{2-6}
 & Follower count & Left/anti & 2.68 & 0.0073 &  0.029\\
%\cline{2-6}
\hline
 & Tweet count & Right/pro & 53.77 & $<$1e-20 & $<$1e-20 \\
%\cline{2-6}
 & Tweet count & Right/anti & 85.93 & $<$1e-20 &  $<$1e-20\\
%\cline{2-6}
 & Tweet count & Left/pro & 53.33 & $<$1e-20 &  $<$1e-20\\
%\cline{2-6}
\multirow{8}{4em}{December 2018} & Tweet count & Left/anti & 31.29 & $<$1e-20 & $<$1e-20 \\
%\cline{2-6}
& Follower count & Right/pro & 13.31 & $<$1e-20 & $<$1e-20 \\
%\cline{2-6}
 & Follower count & Right/anti & 3.72 & 0.0002 &  0.0008 \\
%\cline{2-6}
& Follower count & Left/pro & 1.87 & 0.062 & 0.25  \\
%\cline{2-6}
 & Follower count & Left/anti & 2.71 & 0.0066 &  0.026 \\
\bottomrule
    \end{tabular}
    \caption{Un-adjusted and adjusted (using Bonferroni correction) p-values for each unpaired Welch's t-test.  Users in each political group were compared to users without any political identities in their profile.}
    \label{tab:t_tests}
\end{table}

\section{Average and variance in alignments by category.}

Table 11
%\ref{tab:lr_sum} 
and Table 12
%\ref{tab:est_sum} 
contains the average and standard deviation of alignments along the left/right and pro-/anti-establishment dimension, respectively.

\begin{table}[h]
\begin{tabular}{@{}lccccc}
\toprule
Category &  September 2016 & June 2017 & December 2017 & June 2018 & December 2018 \\
\midrule
sports &  -0.014 (0.41) & -0.0086 (0.42) & -0.037 (0.41) & -0.037 (0.42) &-0.055 (0.42) \\
\midrule
occupation/industry &  -0.26 (0.66) & -0.28 (0.66) & -0.29 (0.66) & -0.32 (0.66) & -0.33 (0.66)  \\
\midrule
school/subjects &  -0.31 (0.73) & -0.32 (0.75) & -0.34 (0.76) & -0.37 (0.76) & -0.37 (0.77)  \\
\midrule
age &  -0.14 (0.28) & -0.19 (0.29) & -0.19 (0.29) & -0.23 (0.3) & -0.25 (0.31) \\
\midrule
culture & -0.46 (0.77) & -0.45 (0.74) & -0.45 (0.74) & -0.46 (0.74) & -0.46 (0.73) \\
\midrule
technology/gaming/cars &  -0.11 (0.44) & -0.11 (0.45) & -0.13 (0.45) & -0.15 (0.44) & -0.18 (0.44) \\
\midrule 
nationality/location/ethnicity/race &  -0.17 (0.83) & -0.17 (0.86) & -0.19 (0.87) & -0.21 (0.86) & -0.22 (0.85) \\
\midrule
family &  0.34 (1.034) & 0.33 (1.007) & 0.31 (0.99) & 0.29 (0.96) & 0.28 (0.97) \\
\midrule
outdoors/nature/animals &  -0.14 (0.89) & -0.16 (0.87) & -0.17 (0.86) & -0.18 (0.85) & -0.18 (0.85) \\
\midrule
relationship status &  -0.31 (0.29) & -0.30 (0.26) & -0.31 (0.34) & -0.30 (0.41) & -0.27 (0.46) \\
\midrule
disability/(mental) illness &  -0.41 (0.59) & -0.40 (0.57) & -0.43 (0.58) & -0.45 (0.56) & -0.48 (0.55) \\
\midrule
religion &  0.34 (1.95) & 0.36 (1.99) & 0.34 (1.95) & 0.32 (1.93) & 0.32 (1.90)  \\
\midrule 
sexuality/gender & -0.40 (1.052) & -0.42 (1.065) & -0.46 (1.045) & -0.51 (1.042) & -0.57 (1.069) \\
\midrule 
travel &  -0.63 (0.52) & -0.58 (0.51) & -0.61 (0.55) & -0.63 (0.53) & -0.63 (0.53)  \\
\midrule
(social) media &  -0.13 (0.71) & -0.099 (0.71) & -0.11 (0.70) & -0.15 (0.65) & -0.17 (0.64) \\
\midrule
business/finance &  0.035 (0.77) & -0.0065 (0.73) & -0.048 (0.70) & -0.098 (0.68) & -0.13 (0.66) \\
\midrule
military &  1.85 (1.39) &  1.91 (1.42) & 1.90 (1.48) & 1.93 (1.48) & 1.92 (1.48) \\
\midrule
activism &  -0.63 (2.46) & -0.81 (2.64) & -0.79 (2.70) & -0.74 (2.82) & -0.71 (2.86) \\
\midrule
horoscope &  -0.38 (0.32) & -0.37 (0.39) & -0.35 (0.41) & -0.32 (0.45) & -0.32 (0.48) \\
\midrule
sex &  -0.23 (0.43) & -0.20 (0.47) & -0.20 (0.51) & -0.20 (0.59) & -0.19 (0.59) \\
\midrule
values &  0.75 (2.59) & 0.73 (2.55) & 0.70 (2.54) & 0.64 (2.52) & 0.68 (2.56)  \\
\bottomrule
\end{tabular}
\label{tab:lr_sum}
\caption{Average (standard deviation) of alignments along left/right dimension per category and time period.}
\end{table}

\begin{table}[h]
\begin{tabular}{@{}lccccc}
\toprule
Category &  September 2016 & June 2017 & December 2017 & June 2018 & December 2018 \\
\midrule
sports &  -0.12 (0.50) &  -0.11 (0.47) & -0.10 (0.46) & -0.10 (0.44) & -0.085 (0.46)  \\
\midrule
occupation/industry & -0.18 (0.77) &  -0.14 (0.76) & -0.10 (0.74) & -0.07 (0.71) & -0.05 (0.70)  \\
\midrule
school/subjects &  0.041 (0.57) &  0.088 (0.56) & 0.13 (0.57) & 0.16 (0.58) & 0.18 (0.58)  \\
\midrule
age &  -0.31 (0.55) &  -0.19 (0.59) & -0.15 (0.62) & -0.032 (0.68) & -0.015 (0.67) \\
\midrule
culture &  -0.33 (0.95) &  -0.28 (0.94) & -0.25 (0.92) & -0.21 (0.89) & -0.17 (0.86)  \\
\midrule
technology/gaming/cars &  -0.41 (0.55) &  -0.37 (0.51) & -0.31 (0.48) & -0.29 (0.45) & -0.26 (0.43) \\
\midrule
nationality/location/ethnicity/race & -0.13 (0.71) &  -0.11 (0.77) & -0.076 (0.79) & -0.043 (0.81) & -0.025 (0.80)  \\
\midrule
family &  0.64 (1.29) &  0.61 (1.28) & 0.56 (1.23) & 0.50 (1.18) & 0.43 (1.17) \\
\midrule
outdoors/nature/animals & 0.16 (0.58) &  0.19 (0.56) & 0.18 (0.54) & 0.17 (0.52) & 0.19 (0.54)  \\
\midrule
relationship status &  -0.14 (0.44) &  -0.053 (0.51) & -0.088 (0.55) & -0.067 (0.59) & -0.065 (0.64)  \\
\midrule
disability/(mental) illness &  0.083 (0.52) &  0.053 (0.56) & 0.077 (0.53) & 0.11 (0.57) & 0.087 (0.58)  \\
\midrule
religion &  0.51 (1.64) &  0.45 (1.67) & 0.43 (1.61) & 0.39 (1.60) & 0.35 (1.60)  \\
\midrule
sexuality/gender &  -0.18 (0.97) &  -0.20 (1.15) & -0.17 (1.19) & -0.12 (1.21) & -0.12 (1.24)  \\
\midrule
travel &  -0.024 (0.38) &  0.013 (0.37) & 0.11 (0.38) & 0.13 (0.36) &  0.13 (0.37) \\
\midrule
(social) media &  0.02 (0.76) & -0.064 (0.79) & -0.031 (0.80) & 0.023 (0.81) &  0.064 (0.80)  \\
\midrule
business/finance &  -0.32 (0.84) & -0.27 (0.87) & -0.20 (0.87) & -0.13 (0.86) &  -0.11 (0.83)  \\
\midrule
military &   0.98 (1.0) & 0.68 (1.029) & 0.40 (1.012) & 0.087 (1.049) & -0.073 (1.049) \\
\midrule
activism & -0.18 (1.099) & -0.22 (1.072) & -0.26 (1.24) & -0.32 (1.55) & -0.34 (1.76) \\
\midrule
horoscope &  0.032 (0.23) & 0.053 (0.19) & 0.060 (0.21) & 0.026 (0.26) & 0.066 (0.31)  \\
\midrule
sex & -0.21 (0.27) & -0.29 (0.37) & -0.30 (0.49) & -0.31 (0.66) & -0.25 (0.68) \\
\midrule
values &  0.21 (1.15) & 0.13 (0.82) & 0.11 (0.57) & -0.064 (0.54) & -0.19 (0.76)  \\
\bottomrule
\end{tabular}
\label{tab:est_sum}
\caption{Average (standard deviation) of alignments along pro-/anti-establishment dimension per category and time period.}
\end{table}




\end{appendices}


\bigskip

\bibliography{biblio}% common bib file

\end{document}